\documentclass[showpacs, pre, superscriptaddress,nofootinbib, twocolumn]{revtex4}

\usepackage{graphicx}
\usepackage{color}
\usepackage{amssymb,amsfonts,amsmath}
\usepackage{natbib}
\usepackage{epstopdf}
\usepackage{subfigure}
\usepackage{braket}
\usepackage{lipsum}

\definecolor{orange}{rgb}{1,0.5,0}

\usepackage{adjustbox}
\usepackage{subfigure}
\usepackage{rotating}

\begin{document}

\title{Inferring network structure and local dynamics from neuronal patterns with quenched disorder}

\author{Ihusan Adam}
\affiliation{Dipartimento di Ingegneria dell'Informazione, Universit\`{a} degli Studi di Firenze, Via S. Marta 3, 50139 Florence, Italy}

\author{Gloria Cecchini}
\affiliation{Dipartimento di Fisica e Astronomia and CSDC, Universit\`{a} degli Studi di Firenze, via G. Sansone 1, 50019 Sesto Fiorentino, Italia}

\author{Duccio Fanelli}
\affiliation{Dipartimento di Fisica e Astronomia and CSDC, Universit\`{a} degli Studi di Firenze, via G. Sansone 1, 50019 Sesto Fiorentino, Italia}
\affiliation{INFN Sezione di Firenze, via G. Sansone 1, 50019 Sesto Fiorentino, Italia}

\author{Thomas Kreuz}
\affiliation{Institute for Complex Systems, CNR, Sesto Fiorentino, Italy}

\author{Roberto Livi}
\affiliation{Dipartimento di Fisica e Astronomia and CSDC, Universit\`{a} degli Studi di Firenze, via G. Sansone 1, 50019 Sesto Fiorentino, Italia}
\affiliation{INFN Sezione di Firenze, via G. Sansone 1, 50019 Sesto Fiorentino, Italia}

\author{Matteo di Volo}
\affiliation{Unit\'e de Neuroscience, Information et Complexit\'e (UNIC), CNRS FRE 3693, 1 avenue de la Terrasse, 91198 Gif sur Yvette, France}

\author{Anna Letizia Allegra Mascaro}
\affiliation{Neuroscience Institute, CNR, Pisa, Italy}
\affiliation{LENS, Universit\`{a} degli Studi di Firenze, via N. Carrara 1, 50019 Sesto Fiorentino, Italia}

\author{Emilia Conti}
\affiliation{Dipartimento di Fisica e Astronomia and CSDC, Universit\`{a} degli Studi di Firenze, via G. Sansone 1, 50019 Sesto Fiorentino, Italia}
\affiliation{LENS, Universit\`{a} degli Studi di Firenze, via N. Carrara 1, 50019 Sesto Fiorentino, Italia}

\author{Alessandro Scaglione}
\affiliation{Dipartimento di Fisica e Astronomia and CSDC, Universit\`{a} degli Studi di Firenze, via G. Sansone 1, 50019 Sesto Fiorentino, Italia}
\affiliation{LENS, Universit\`{a} degli Studi di Firenze, via N. Carrara 1, 50019 Sesto Fiorentino, Italia}

\author{Ludovico Silvestri}
\affiliation{Dipartimento di Fisica e Astronomia and CSDC, Universit\`{a} degli Studi di Firenze, via G. Sansone 1, 50019 Sesto Fiorentino, Italia}
\affiliation{LENS, Universit\`{a} degli Studi di Firenze, via N. Carrara 1, 50019 Sesto Fiorentino, Italia}

\author{Francesco Saverio Pavone}
\affiliation{Dipartimento di Fisica e Astronomia and CSDC, Universit\`{a} degli Studi di Firenze, via G. Sansone 1, 50019 Sesto Fiorentino, Italia}
\affiliation{LENS, Universit\`{a} degli Studi di Firenze, via N. Carrara 1, 50019 Sesto Fiorentino, Italia}
\affiliation{Istituto Nazionale di Ottica (INO), National Research Council, Via Nello Carrara 1, 50019 Sesto Fiorentino, Italy}

% Keywords are not mandatory, but authors are strongly encouraged to provide them. If provided, please include two to five keywords, separated by the pipe symbol, e.g:
%\keywords{ Network $|$ Inverse problem $|$ Neuronal dynamics $|$ Leak Integrate and Fire $|$ Heterogenous Mean Field $|$ Animal stroke models $|$  Fluorescence microscopy} 

\begin{abstract}
An inverse procedure is proposed and tested which aims at recovering the a priori unknown functional and structural information from global signals of living brains activity. To this end we consider a 
Leaky-Integrate and  Fire (LIF) model with short term plasticity neurons, coupled via a directed network.  Neurons are assigned a specific current value, which is heterogenous across the sample, and sets the firing regime in which the neuron is operating in. The aim of the method is to recover the distribution of  incoming network degrees, as well as the distribution of the assigned currents, from global field measurements. The proposed approach to the inverse problem 
implements the reductionist Heterogenous Mean-Field approximation. This amounts in turn to  organizing the neurons in different classes, depending on their associated degree and current.  When tested again synthetic data,  
the method returns accurate estimates of the sought distributions, while managing to reproduce and interpolate almost exactly the time series of the supplied global field.  Finally, we also applied the proposed technique to 
longitudinal wide-field fluorescence microscopy data of cortical functionality in groups of awake Thy1-GCaMP6f mice. Mice are induced a photothrombotic stroke in the primary motor cortex and their recovery monitored in time. An all-to-all
LIF model which accommodates for currents heterogeneity allows to adequately explain the recorded patterns of activation. Altered distributions in neuron excitability are in particular detected,  compatible with the phenomenon of  
hyperexcitability in the penumbra region after stroke.
\end{abstract}

\maketitle
%\thispagestyle{firststyle}
%\ifthenelse{\boolean{shortarticle}}{\ifthenelse{\boolean{singlecolumn}}{\abscontentformatted}{\abscontent}}{}

Living brains display extraordinarily complex and rich dynamical behaviors, often modeled by resorting to  
ensembles made of simple biological neurons, coupled via an intricate web of interlinked connections.  These latter define the 
architecture of the embedding network. The computational power of a  neuronal system stems from the peculiar 
non-linear features, as displayed by individual units, and their mutual interactions, as mediated by the 
network topology. Inferring network characteristics from direct measurements of neuronal activity (spiking events), 
constitutes a challenging task, which yielded a large plethora of different methodological approaches. Popular
techniques, aimed at recovering quantitative information on existing inter-nodes connections, rely for instance on 
statistical mechanics concepts, as e.g. maximum entropy principles \cite{bialek,cocco}.
Network topology is  one out of many possible sources of heterogeneity in the brain. Neurons may in fact also exhibit 
different intrinsic dynamics, an additional ingredient which can significantly impact the functioning of the brain \cite{het_coding}. Neuronal excitability 
(i.e. neurons' ability to respond to external inputs) is finely orchestrated in the brain through inhibitory/excitatory balance \cite{deghani, neuromod}. Moreover, dysregulation of neurons' effective excitability (e.g. through excitation/inhibition unbalance) is often cause of pathologies, for example epilepsy \cite{balance_epilepsy}. Stroke is in particular strongly associated with alteration in excitatory/inhibitory balance whose recovery can be boosted by rehabilitation \cite{AllegraMascaro,spalletti1}. These factors are therefore crucial to understand either the onset or the evolution of pathological states. Motivated by this, we here propose and successfully test against both synthetic and real data, 
an inverse protocol scheme which is targeted to quantifying the  neurons' inherent (quenched) excitability, while inferring the, a priori unknown, distribution of network connectivities, from global activity measurements.  

To this end we consider an extended model of Leaky-Integrate and 
Fire (LIF) neurons,  with short-term plasticity. The neurons are coupled to 
a directed network and display a degree of heterogeneity in the associated 
current, which sets the degree of effective excitability and thus the firing regime in which a neuron is forced to operate. 
The aim of the  method is to recover the distribution of the (in-degree) connectivity, in the following
labelled $k$, which characterizes the embedding network, as well as the distribution of the assigned currents, denoted by $a$.

Our approach to the inverse problem builds on the celebrated Heterogenous 
Mean-Field (HMF) approximation. The key idea is  to rewrite the dynamics of the system by organizing  
the neurons in different classes, each associated to distinct values of the
current $a$ and of the connectivity $k$. The HMF reduction scheme allows us to create a 
mesh in the space of the variables $a$ and $k$, in such a way that the contribution of all possible  
neurons can be adequately represented. The inverse scheme aims at determining the sought density distributions  
$P(a)$ and $P(k)$, as iterative solutions of a Fredholm-like equation which self-consistently identifies the classes of 
neurons needed to confidently reproduce the global activity field, supplied as an input to the algorithm. 
The HMF approximation has been previously employed in \cite{diVolo1,diVolo2,diVolo3, AFCI}  to recover the information on the 
hidden network topology, from global and local, synthetically generated, data. Starting from these premises, we here take a decisive leap forward by proving 
that a non trivial generalization of the method can be effectively invoked to quantify, in addition,  the amount of dynamical heterogeneity that characterizes the examined 
process. As we shall argue in the following, this latter component exerts a role of paramount importance in driving non periodic, seemingly irregular patterns of activity, 
like those displayed in real measurements. Our results are manyfold. First, we will introduce the inverse method and test its    
efficacy against synthetic data. These are generated by assuming a random network to provide the structural skeleton of the model, while at the same time  imposing a 
single- or multi-peaked profile for the distribution of currents. The method allows to accurately estimate the 
two distributions, while managing to interpolate almost exactly the time series of the collected global field. Then, we build a bridge from measurements to theory, by arguing that the raster plots for the spiking activity of  
neurons over time can provide the ideal input for the reconstruction method to work. Finally, we apply the proposed technique to experimental data, in this case
longitudinal wide-field fluorescence microscopy data of cortical functionality in groups of awake mice. 
To mimic an ischemic condition, a phototrombotic stroke is induced in the primary motor cortex. Functional recovery is promoted, via motor training on a robotic platform.
The global activity fields as obtained from the experiments can be nicely interpolated by a LIF model which accommodates for quenched heterogeneity in the currents.  Altered  distributions in neuron excitability are detected 
during the acute phase, i.e. immediately after stroke. In particular, the neurons in the region adjacent to the stroke manifest a degree of enhanced excitability, in agreement with earlier experimental observation \cite{HyperExcitability1,HyperExcitability2,HyperExcitability3}.  Conversely, rehabilitation recovered a distribution more similar to pre-stroke conditions. Taking all together, our results suggest that the proposed method could serve as a useful tool to
follow the recovery process from a stroke. 

\section*{Results}

{\bf The model.} We consider a network of $N$ nodes, labelled with the progressive discrete index $i$. Each node identifies a LIF neuron.  Neurons have equal synapses and interact via the synaptic currents 
that are regulated by short-term plasticity.  This model was proved capable of reproducing a large variety of dynamical regimes, from a-synchronous  to quasi-synchronous regimes, encompassing bursting and 
chaotic behaviors \cite{pittorino,TM,diVolo1}.  Each neuron is associated a dynamical state vector ${\bf w}_i(t) = (v_i(t), x_i(t), y_i(t), z_i(t))$,
 where $v_i$ stands for the rescaled membrane potential and $x_i$, $y_i$ and $z_i$ denote the fractions of neurotransmitters in the available, active and inactive states,
respectively. This implies that, at any time $t$
\begin{equation}
x_i(t)+ y_i(t) +z_i(t) = 1
\label{norma}
\end{equation}
The dynamics of the neural network is ruled by the following equations
\begin{subequations}
\begin{equation}
{\dot v}_i(t) = a_i - v_i(t) + \frac{g}{N} \sum_{j \not= i} A_{ij} \, y_j(t)
\label{LIFmodel1}
\end{equation}
\begin{equation}
{\dot y}_i(t) = - \frac{y_i(t)}{\tau_{in}} + u x_i(t) S_i(t)
\label{LIFmodel2}
\end{equation}
\begin{equation}
{\dot z}_i(t) = \frac{y_i(t)}{\tau_{in}} - \frac{z_i(t)}{\tau_r}
\label{LIFmodel3}
\end{equation}
\end{subequations}
where $a_i$ represents the external input current of the $i$-th neuron. As mentioned above,
this is an important ingredient of the model, as it regulates the endogenous functioning of the corresponding neurons, in the uncoupled limit. 
With reference to the adopted rescaled variables setting, the critical value $a_i=1$ identifies 
a bifurcation point, which separates between quiescent and  active (spiking) dynamical regimes.
Notwithstanding the focus of the model that is here solely placed on excitatory neurons (i.e. $g>0$), 
$a$ can be conceptualized as an effective parameter which controls the net current received by the neuron, forcing it to
input ($a>1$) or fluctuation ($a<1$) dominated  regimes.  We anticipate that the forthcoming analysis generalizes  to the interesting setting where 
populations of excitatory and inhibitory neurons are simultaneously present, at the price of a non trivial complexification of the procedures involved \cite{Chicchi}.   

In Eq. (\ref{LIFmodel2}), $S_i(t)$ represents the spike train produced by neuron $i$, in 
formulae
\begin{equation}
S_i(t) = \sum_n \delta(t-t_i(n))
\label{spiketrain}
\end{equation}
where $t_i(n)$ is the time when neuron $i$ fires its $n$-th spike. Notice that, in this simplified model of neural
activity, a spike is assimilated to a Dirac $\delta$-distribution. Such a spike is generated by neuron $i$ whenever its
membrane  potential $v_i(t)$ crosses the threshold value $v_{th} = 1$. Immediately afterwards  
$v_i$ is reset to zero and the spike is transmitted to its efferent neurons. The synaptic dynamics regulated by 
the short-term plasticity which mediates the transmission of the spike-train function, follows the scheme implemented in  
Eqs. (\ref{LIFmodel2}) and (\ref{LIFmodel3}). When neuron $i$ emits a spike, it releases a fraction of neurotransmitters 
$u x_i(t)$ and the fraction of active resources $y_i(t)$ gets consequently increased. In between consecutive spikes of 
neuron $i$, the use of active resources yields an exponential 
decrease, on a time scale $\tau_{in}$, of $y_i(t)$, while the fraction of inactive resources $z_i(t)$ gets increased by the same 
amount. At the same time, while $z_i(t)$ decreases, over a time scale $\tau_r$, the fraction of available resources is eventually reintegrated. 
In fact, by combining Eqs. (\ref{LIFmodel2}),  (\ref{LIFmodel3}) and (\ref{norma}) one readily obtains
\begin{equation}
{\dot x}_i(t) = \frac{z_i(t)}{\tau_r} - u x_i(t) S_i(t).
\label{avres}
\end{equation}
Spikes from the afferent neurons $j$ are received by neuron $i$ through the sudden increase of their active resources
$y_j(t)$ and affect the dynamics of its membrane potential via the coupling term (tuned by the parameter $g$) which sits on
the right hand side of Eq. (\ref{LIFmodel1}). The existing connections between pairs of neurons are specified by the binary adjacency matrix ${\bf A}$. 
The entries $A_{ij}$ are set to one if neuron $j$ fires to neuron $i$ and zero, otherwise. In this simplified model, all the parameters appearing in the above equations (except for the intensity parameter $a_i$) 
are independent of the neuron indices  and each neuron is connected to a macroscopic number of pre-synaptic neurons: this is the reason why the interaction 
coupling $g$ is normalized by the factor $N$. We further consider {\sl dense networks}, where 
each neuron is connected to ${\mathcal O}(N)$ pre-synaptic neurons \footnote{This choice is inspired to the
the mean-field derivation of the HMF technique, the standard mean-field limit being recovered
in a fully connected network, where $A_{ij} = 1$ for any pair $i,j$, i.e. where the number of pre-synaptic neurons is
$N-1$ for all neurons. We stress however to that the HMF works out also for 
non-dense networks.}. As it is common practice, the timescales in the above equations are attributed  by the phenomenological
values in adimensional units, $\tau_{in} = 0.2$, $\tau_r = 133 \tau_{in}$, $g = 30$ and $u = 0.5$, see \cite{DLLPT,TM}.

{\bf The heterogeneous mean field ansatz.}
A simple analogue of the model described in the previous Section, with homogeneous currents, $a_i = a >1$, was studied in 
\cite{diVolo1}, and successfully employed to reconstruct in silico the network structure (more precisely, the distribution function from which the 
random connectivity network was sampled) from global signal of, synthetically generated, synaptic activity. To accomplish this step, the 
original model is reformulated in terms of a Heterogeneous Mean Field (HMF) approximation:
the original single-neuron variables are hence replaced by families of neurons with equal average input connectivity. In practice,
when dealing with an extended population of neurons, connected via a large number of synapses, one can safely enough omit unimportant 
{\it microscopic} details of the network structure and just focus on average, i.e. mean-field-like, observables. The goal of our analysis is to take the 
method to a different conceptual level, by allowing for a quenched disorder in the current input.  Quenched disorder turns out to be in a crucial factor for 
capturing the variability of the amplitude of the global signal of synaptic activity,  which appears to be unnaturally regular in the homogeneous case, 
i.e. when all $a_i=a$, as in \cite{diVolo1}. {\sl A posteriori} we can argue that quenched disorder amounts to a heuristic recipe for surrogating the effects of inhibitory neurons, 
in a simplified framework where neurons are assumed of the excitatory type. To work out the HMF approximation for the considered model we define the 
the input field $Y_i$ to each neuron as $Y_i=\frac{1}{N}\sum_j A_{ij} y_j$. This latter quantity can be written in terms of the average input field per neuron as 
$Y_i=\frac{k_i}{N}\left[\frac{1}{k_i}\sum_{j \in I(i)} y_j \right]$ where $I(i)$ represents the set of $k_i$ pre-synaptic neurons of neuron $i$. 
For large enough networks, the average input field for neuron $i$ can thus be approximated as $\frac{1}{k_i}\sum_{j \in I(i)} y_j  \approx \frac{1}{N}\sum_{j} y_j$.
Since $Y(t)=\frac{1}{N}\sum_{j} y_j$ and that $\tilde{k}=\frac{k_i}{N}$, we can write $Y_i=\tilde{k}Y(t)$ where $\tilde{k}\in\left(0,1\right]$. 
The key idea is to assume that all neurons with the identical rescaled degree $\tilde{k}$ and current $a$ share the same dynamical equations, a condition which readily translates 
into the self-consistent dynamical relation  

\begin{equation}
\label{eq10}
Y(t)=\int_0^1\int_{a_{min}}^{a_{max}}P(\tilde{k})P(a) y_{\tilde{k},a}  da dk
\end{equation}

where $a\in (a_{min},a_{max})$ and

\begin{equation}
\label{eq11}
\dot{v}_{\tilde{k},a}= a- v_{\tilde{k},a}(t) + g \tilde{k} Y(t)
\end{equation}
\begin{equation}
\label{eq12}
\dot{y}_{\tilde{k},a}= -\frac{y_{\tilde{k},a}(t)}{\tau_{in}} + u\left(1-y_{\tilde{k},a}(t) - z_{\tilde{k},a}(t)\right)    S_{\tilde{k},a}(t)
\end{equation}
\begin{equation}
\label{eq13}
\dot{z}_{\tilde{k},a}= \frac{y_{\tilde{k},a}(t)}{\tau_{in}} - \frac{z_{\tilde{k},a}}{\tau_r}
\end{equation}

The above set of equations provides a great numerical advantage over their original analogues, especially when large systems are to be considered. 
Each class of neurons, as identified by the combined topological/dynamical pair $(\tilde{k}, a)$, is in fact solely driven by the mean-field $Y(t)$. It is hence possible to 
update in time the state of the system in the reduced space of the reference classes without keeping explicit track of all binary interactions, as encoded in the network adjacency matrix.
Further, the basic features of the dynamics of the examined system can be effectively reproduced (modulo finite--size corrections) by exploiting a suitable discrete sampling of both 
$P(\tilde{k})$ and $P(a)$ (see below). The validity of the HMF ansatz is challenged via direct simulations of the original model. Results reported in annexed Supplementary Information (SI) testify on the adequacy of the 
proposed approximation.

{\bf The inversion method}
The input to the inverse problem is represented by the global field $Y(t)$, which is therefore assumed known. We shall return to the actual generation of the time series $Y(t)$, 
when discussing the application of the proposed method to both synthetic and real data. In the following, we will set up the general scheme for recovering the unknown distributions 
$P(a)$ and $P(\tilde{k})$, by interpolating the available field  $Y(t)$, under the simplified HMF descriptive framework. 

First of all, and without any loss of generality, we select a suitable interval $[a_{min},a_{max}]$, which contains the bifurcation threshold $a=1$ and provides the support for the function $P(a)$ that we aim at recovering. 
The choice of the interval width is completely arbitrary and just impacts the ensuing analysis in term of associated computational cost: the larger the interval, the more demanding the numerical charge.
We then proceed by discretizing the interval in $q$ equally spaced bins, each of extension $\Delta a = (a_{max}-a_{min})/q$. The discrete counterpart of the continuous distribution $P(a)$ is therefore traced back to a
vector made of $q$ scalar unknown components, namely ${\bf P_a}=(P(a_1)\cdots P(a_q))$.

We proceed similarly to discretize the continuous function $P(\tilde{k})$. The unitary segment is binned in $r$ equally spaced intervals, each of size $\Delta \tilde{k}=1/r$,  to eventually  obtain 
the $r$-dimensional vector ${\bf P_{\tilde{k}}}=(P(\tilde{k}_1) \cdots P(\tilde{k}_r))$, which constitutes the discrete version of the degree distribution that we seek at reconstructing. 

The inversion procedure runs as follows. The HMF model made of Eqs. (\ref{eq11}),(\ref{eq12}) and (\ref{eq13}) is initialized, with the variables ($v_{\tilde{k},a},y_{\tilde{k},a},z_{\tilde{k},a}$) being randomly drawn from a uniform distribution, which  is constructed so as to respect the obvious constraints $v_{\tilde{k},a}<1$ and $y_{\tilde{k},a}+z_{\tilde{k},a}<1$. The governing equations, forced by the external field $Y(t)$, are integrated forward and the variables $y_{\tilde{k},a}$ stored, for each choice of the pair $(\tilde{k},a)$ and for any sampling time. The process is repeated for $M$ independent realizations of the initial conditions and the average fraction of neurotransmitters in the active state, for classes ($\tilde{k},a$)  computed at any time of observation $t_j$, $j=1,...,n$, namely $\langle y_{\tilde{k},a} (t_j) \rangle=1/M \sum_{i=1}^M \left(y_{\tilde{k},a}(t_j)\right)_i$, where the index $i$ identifies distinct trajectories. Our aim is to determine the (normalized) vectors  
${\bf P_a}$ and ${\bf P_{\tilde{k}}}$ which match the self-consistent condition $Y(t_j) = \sum_{l=1}^q \sum_{m=1}^r P_{a_l} P_{\tilde{k}_m} \langle y_{\tilde{k},a} (t_j) \rangle$, for any time $t_j$. To this end, we start with an initial guess for the distribution to be eventually recovered (typically a uniform distribution over the respective interval of pertinence) and iterate a recursive scheme which is organized into two nested cycles.

\begin{figure*}[htb!]
    \centering
    \includegraphics[scale=0.35]{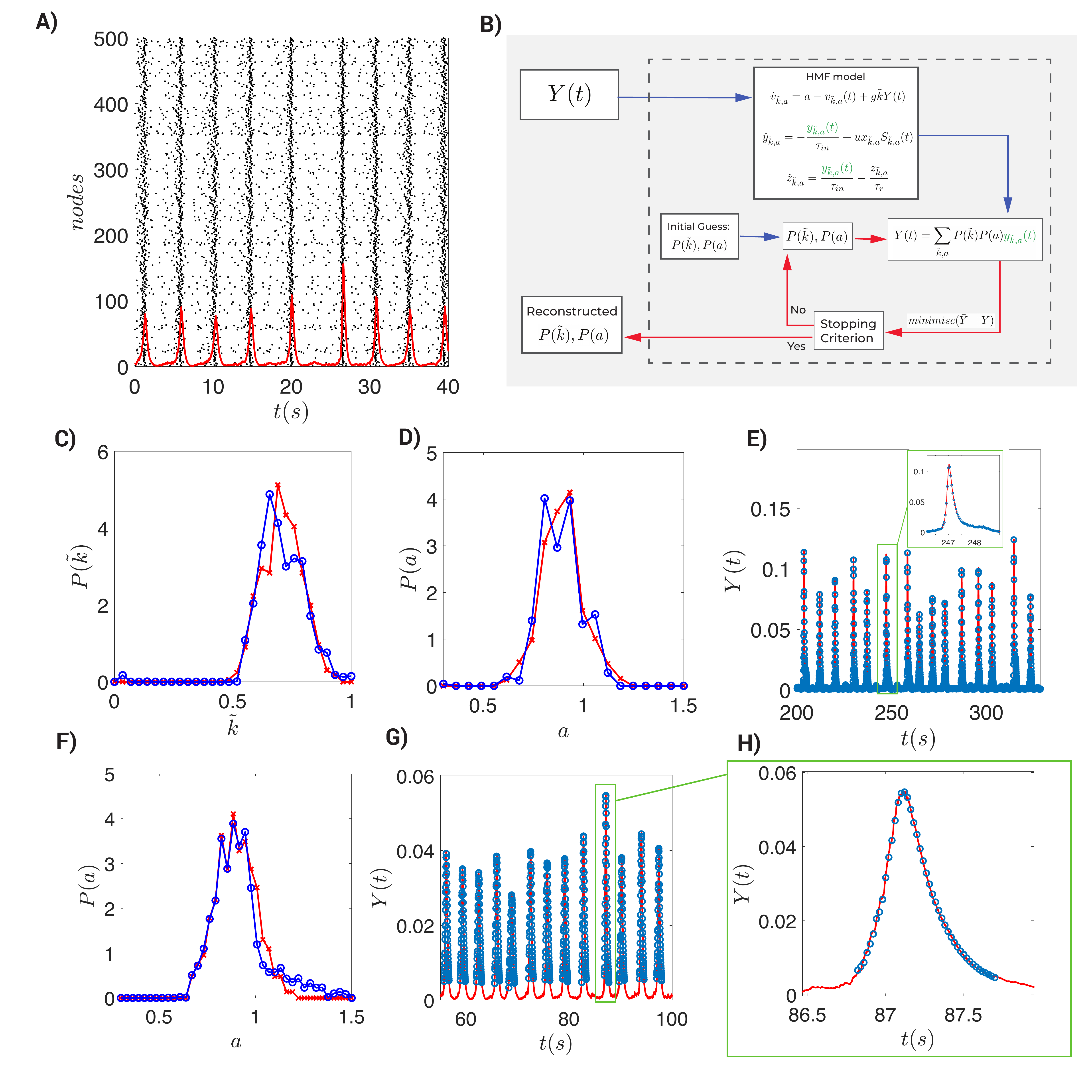}
    \caption{Panel A) shows the raster plot as obtained by simulating a network of $N=500$ nodes. The distributions of in-degree and currents are drawn from Gaussian distributions with, respectively, $\langle k \rangle = 0.7$, $\sigma_{\Tilde{k}}=0.082$ and $\langle a \rangle = 0.9$,  $\sigma_{a}=0.1$).  The global field $Y(t) $ is overlaid in red and multiplied by a scaling factor equal 2500 to help visualization. Panel B) shows a schematic outline of the reconstruction procedure. The blue arrows stand for the flow of inputs while red arrows refer to the iterative minimization procedure. Panels C), D) and E) display the outcome of the procedure, for the setting referenced to in the description of panel A). C) The true in-degree distribution is plotted in red while the recovered distribution is depicted in blue. In panel D) the true distribution of the external currents is displayed  in red and the recovered distribution plotted in blue. E) The true global field ($Y(t)$) is shown with the solid red line and the reconstructed homologue represented with blue circles. The plot in the inset is a zoom in of the peak in the green box. Panels F), G) and H) reports results of a similar test where now the network is fully connected and only $P(a)$ is reconstructed. Here the algorithm seek at interpolating the driving field for value of $Y(t)$, above a minimal threshold. In panel F) the true distribution of the external currents ($\langle a \rangle = 0.9$,  $\sigma_{a}=0.1$) is displayed in red and the recovered distribution in blue. In panel G) the true global field is plotted with a solid red line and the blue circles refer to the reconstructed $Y(t)$. Finally, panel H) shows a zoomed in plot of the peak in the green box in panel G). The plots of the distributions in panels C,D,E and F have been binned to have fewer points than the sampling of $\tilde{k}$ and $a$  ($r$ and $q$, respectively) that was used in the reconstruction.
}
\label{fig:1}
\end{figure*}

First, the assigned components of  ${\bf P_a}$  are used to compute the reduced $y_{\tilde{k}} (t_j) =\sum_{l=1}^q  P_{a_l}  \langle y_{\tilde{k},a} (t_j) \rangle$ entries (notice that the angle brackets are deliberately dropped on the left hand side to lighten the notation). These entries can be organized into a  linear problem which implements the above constrain, for a quenched distribution of currents: 

\textsc{\begin{equation}
\begin{bmatrix}
1/\bigtriangleup \tilde{k} \\ Y_1 \\ \vdots \\ Y_n
\end{bmatrix}
=
\begin{bmatrix}
1 & \cdots & 1 \\
y_{\tilde{k}_1}(t_1) & \cdots & y_{\tilde{k}_r}(t_1) \\
\vdots         & \ddots & \vdots \\
y_{ \tilde{k}_1}(t_n) & \cdots  & y_{\tilde{k}_r } (t_n)
\end{bmatrix}
\begin{bmatrix}
P_{\tilde{k}_1}\\
\vdots\\
P_{\tilde{k}_r}
\end{bmatrix}
\label{eq16}
\end{equation}}
where $(Y_1 \cdots Y_n )$ are the $n$ values of the continuous field $Y(t)$ sampled at the observation times $(t_1,t_2....t_n)$. The first equation of the above linear system is added to enforce the normalization of the sought solution. 
By matrix inversion one can in principle obtain an estimate of ${\bf P_{\tilde{k}}}$, at fixed ${\bf P_a}$. In real applications, system (\ref{eq16}) is solved in norm using dedicated optimization tools, as the transfer $n \times r$ matrix is in general ill-conditioned. As a second step in the reconstruction process, we freeze ${\bf P_{\tilde{k}}}$ to the solution obtained above and compute  $y_{a} (t_j) =\sum_{m=1}^r  P_{\tilde{k}_m}  \langle y_{\tilde{k},a} (t_j) \rangle$ (again we dropped, for the ease of notation, the angle bracket). Then the following linear problem is found:

\textsc{\begin{equation}
\begin{bmatrix}
1/\bigtriangleup a \\ Y_1 \\ \vdots \\ Y_n
\end{bmatrix}
=
\begin{bmatrix}
1 & \cdots & 1 \\
y_{a_1}(t_1) & \cdots & y_{a_q}(t_1) \\
\vdots         & \ddots & \vdots \\
y_{ a_1}(t_n) & \cdots  & y_{a_q }(t_n)
\end{bmatrix}
\begin{bmatrix}
P(a_1)\\
\vdots\\
P(a_q)
\end{bmatrix}
\label{eq18}
\end{equation}}

which can be readily solved via apt numerical methods so as to obtain an estimate of ${\bf P_a}$, at fixed ${\bf P_{\tilde{k}}}$. The above procedure is iterated recursively until a maximum number of allowed cycles is reached, or, alternatively,  the stopping criterion is eventually met. The proposed reconstruction algorithm is graphically illustrated in panel B of Fig.  \ref{fig:1}. In the following section, we will test the predictive adequacy of the method, challenging it against synthetically generated 
data. Then, we will turn to the analysis of longitudinal wide-field fluorescence microscopy data cortical functionality in mice.

{\bf Testing the reconstruction algorithm against synthetic data} We shall hereafter proceed by testing the proposed reconstruction protocol against synthetic data. To this end we begin by generating a random graph made of $N$ nodes. This in turn amounts to constructing the binary $N \times N $ adjacency matrix  ${\bf A}$, which sets the links among pairs of adjacent nodes. The procedure is engineered so as to return a bell shaped distribution of connectivities $P(\tilde{k})$. 
The standard deviation of the distribution is sufficiently small so that the tails of the distribution at the boundaries  are by all practical purposes irrelevant. We then add quenched disorder in the input currents $a_i$. The assigned currents display a mono-modal probability distribution $P(a)$, but more complex scenarios can be considered (see SI, where a bimodal $P(a)$ is assumed to hold). 
In panels C and D of Fig. \ref{fig:1}, the distributions $P(\tilde{k})$ and $P(a)$, as obtained via the recipe outlined above, are depicted in red (crosses). These are the exact distributions that we aim at eventually recovering by means of the proposed reconstruction algorithm. Note that the chosen $P(a)$ extends over a finite domain which includes the bifurcation value $a=1$.

As a first step in the analysis we proceed by integrating forward equations (\ref{LIFmodel1}), (\ref{LIFmodel3}) and (\ref{LIFmodel3}), for our specific realization of the network architecture and accounting for the heterogenous collection of current entries $a_i$. From the numerics, one can readily obtain the ensuing raster plot, displayed in  panel A of Fig. \ref{fig:1}: each black dot identifies a spiking event. Neurons are ranked with a progressive discrete index which runs from low to highly connected units. The smooth curve superposed in red stands for the recorded global field $Y(t)$. As an important remark, we notice that the field $Y(t)$ shows irregular oscillations, characterized by a significant modulation in their relative height, as well as a non uniform distribution of the time interval between successive peaks. This is at variance with the case where the currents are set to an identical value (when i.e. the distribution $P(a)$ converges to a Dirac delta). In this latter case, in fact, the oscillations are periodic and the LIF model cannot be straightforwardly invoked to reproduce the complex heterogeneity of patterns, as displayed in the by brain of living entities. The time series $Y(t)$ is provided as an input to the reconstruction algorithm described above, and schematically illustrated in panel B of Fig. \ref{fig:1}. The reconstructed distributions $P(\tilde{k})$ and $P(a)$ are depicted in blue (circles) in panels C and D of Fig. \ref{fig:1} and adhere to their corresponding exact profiles. $P(\tilde{k})$ and $P(a)$  are initially assumed to be uniform in their domain of pertinence. In order to recover the sought distributions, the algorithm seeks to iteratively interpolate the global field $Y(t)$. The reconstructed field obtained at the end of the procedure is
represented in blue (circles) in panel E of Fig. \ref{fig:1}: it is almost indistinguishable from the curve depicted in red, which refers to the recorded input function $Y(t)$. The quality of the fitting can be better appreciated by performing a zoom on just one peak of the collection (see inset of  panel E). As a side remark, we recall that the reaction parameter of the LIF model are  here set to the nominal values, as assumed in the forward simulations. Panel F of Fig. \ref{fig:1} shows the reconstructed $P(a)$, in blue (circles), as obtained from a numerical test which assumes an all--to--all network of inter-neurons corrections. As we shall elaborate in the following, this will prove a setting of interest for properly addressing the analysis of the experimental time series. The exact $P(a)$ is depicted in red and the quality of the reconstruction is remarkably good. In panel G of Fig. \ref{fig:1} the interpolated $Y(t)$ (blue, circles) is confronted with the original signal (red), as stored in direct simulations and the agreement is again excellent (see also the annexed zoom in panel H). In this case, the iterative scheme was set to minimize the distance between the original and the reconstructed $Y(t)$, just for values of the field above a given threshold (notice that the fitted entries, the blue circles, are depicted only for a sufficiently large activity amount). This operative choice is motivated by the need of challenging a protocol that would be suited when dealing with experimental data. The low signal component is in fact in general corrupted by shot noise and should be filtered out in the reconstruction scheme. Summing up, we have here successfully tested the method against synthetic data. We have in particular demonstrated that the proposed scheme is  capable to self-consistently recover the distributions of degrees and currents (assumed independent and not mutually entangled), from a heterogeneous, seemingly irregular, field of neuronal activity.  In the next Section, we will apply the method to  
longitudinal wide-field fluorescence data of cortical functionality in mice.

\begin{figure*}[htb!]
    \centering
    \includegraphics[scale=0.35]{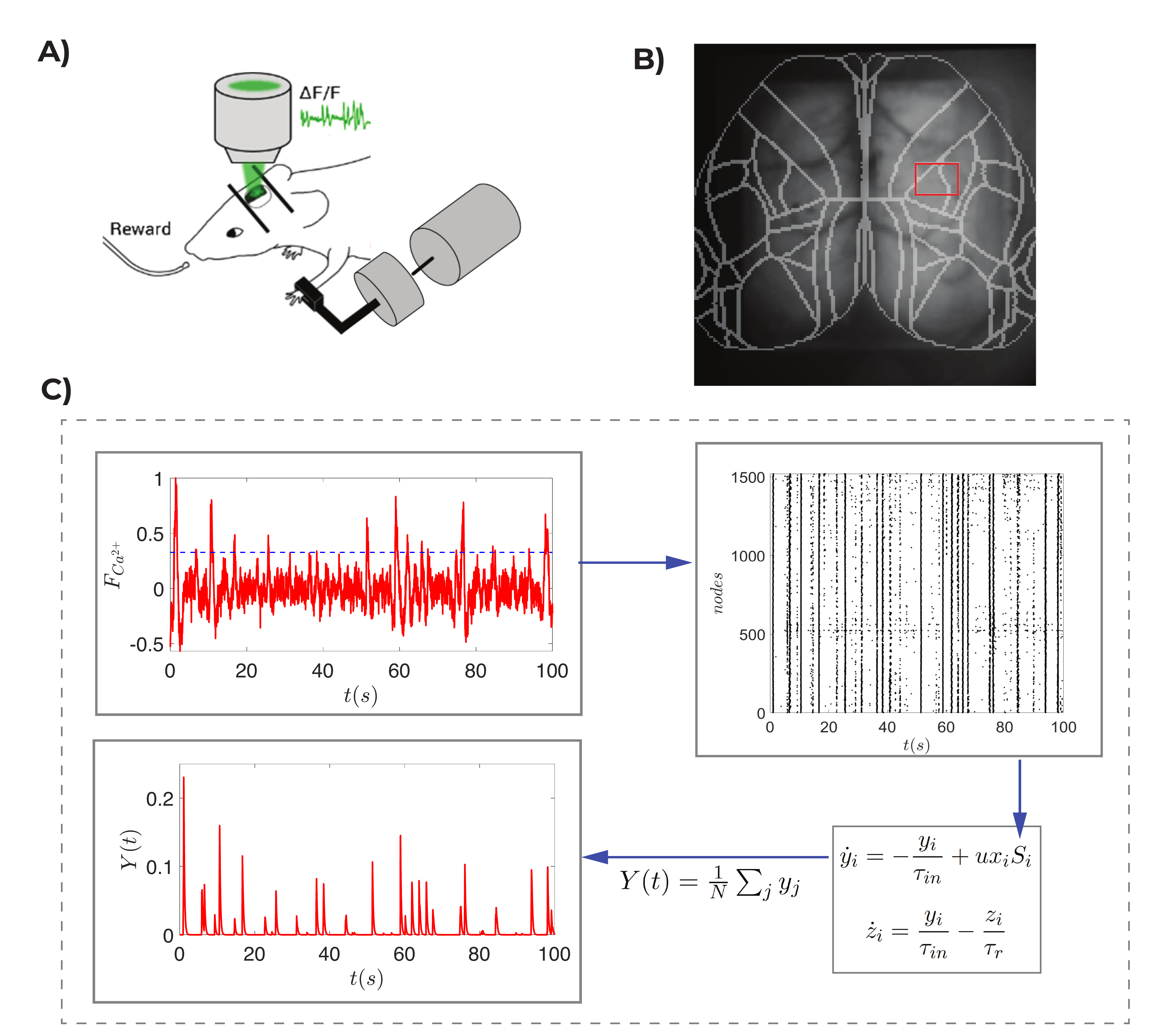}
    \caption{In panel A) a schematic layout of the experimental setup is shown. This includes the imaging equipment and the M-Platform used for rehabilitative training. Panel B) depicts (in red)  the region of the brain from which the data for the analysis are eventually extracted. The region of interest is made of $20 \times 20$ pixels and encloses the portion of the brain affected by the stroke. The patterns in color of grey displayed in the background refers to one typical snapshot of the calcium dynamics. The different domain in which the brain is segmented follow the Allen brain atlas. The first plot in panel C)  shows the calcium florescence signal from one isolated pixel among those which fall in the region of interest, after detrending  performed with a moving running average over a sliding window of three seconds. The to-right plot in panel C) is  the raster plot of detected spikes using a threshold on each pixel. The threshold is here set to $1.8 \sigma$ from the mean, where $\sigma$ stands for the standard deviation of the recorded signal. Next, the spike train information is used to integrate the dynamics on each node, by means of Eqs. (\ref{LIFmodel2}) and (\ref{LIFmodel3}). The experimentally determined raster provides the needed function $S_i(t)$. Form the knowledge of $y_i$ (on all pixels) one can compute the global field $Y(t)$ which is needed as an external driving input of the reconstruction scheme.}
    \label{fig:2}
\end{figure*}

{\bf Analysis of fluorescence data.} The reconstruction technique developed above will be here employed  to quantitatively analyze the wide-field fluorescence microscopy data of  cortical functionality collected from transgenic mice expressing a functional indicator in excitatory neurons. The resting state and motor-evoked cortical activation was recorded before, acutely after stroke and following rehabilitative training.
For a technical account of the experimental setup we refer to the  annexed Method. Starting from the detrended calcium fluorescence images (see panel C of Fig. \ref{fig:2}), for each pixel of the ensemble which defines the region of interest, we construct a raster plot as depicted in panel C of Fig. \ref{fig:2}. To this end, we set a proper threshold and register, for each pixel, the time of exact crossing. The selected events are represented as  black dots in panel C of Fig. \ref{fig:2} and provide a direct evidence for the level of neuronal activity within the associated pixel. It is worth emphasizing that each pixel returns a signal which is the convoluted sum of the fluorescence signal emitted by lots of neurons laying both on the superficial and deep layers of the corresponding brain region. For what concerns the analysis that follows, we shall consider each pixel as representing one individual macroscopic neuron and assume that the interplay between coarse grained units is ruled by a LIF model of the type analyzed above. The distribution of degrees and currents that we shall hereafter recover should be hence conceptualized with reference to the imposed level of spatial abstraction. The raster plot, as obtained from the experimental data, can be straightforwardly employed to compute the global field $Y(t)$, the input of the reconstruction algorithm discussed above. More specifically,  the spike train information is a proxy for the function $S_i(t)$, which in spirit of the self-consistent formulation of the LIF model, comes from the dynamics of  the membrane potential $v_i$. In this case, we take advantage of the experimentally computed $S_i(t)$ ($i$ running on the pixels that define the region of interest) to evolve the paired  Eqs. (\ref{LIFmodel2}) and (\ref{LIFmodel3}). We hence obtain the explicit evolution for the variables $y_i$, which can be combined so as to obtain the  collective field $Y(t)$ (see schematic picture of panel C in Fig. \ref{fig:2}). Notice that the reaction parameters which enter the definition of the above equation can be set to the nominal (although experimentally motivated) values mentioned above. These latter parameters will be also assumed when running the reconstruction scheme. In practical terms,  Eqs. (\ref{LIFmodel2}) and (\ref{LIFmodel3}), complemented with the experimental input which materializes in the raster plot and, hence, in the associated function $S_i(t)$, act as a veritable filter which transforms fluorescence data in an ideal input for the inverse procedure to run.  

\begin{figure*}[htb!]
    \centering
    \includegraphics[scale=0.35]{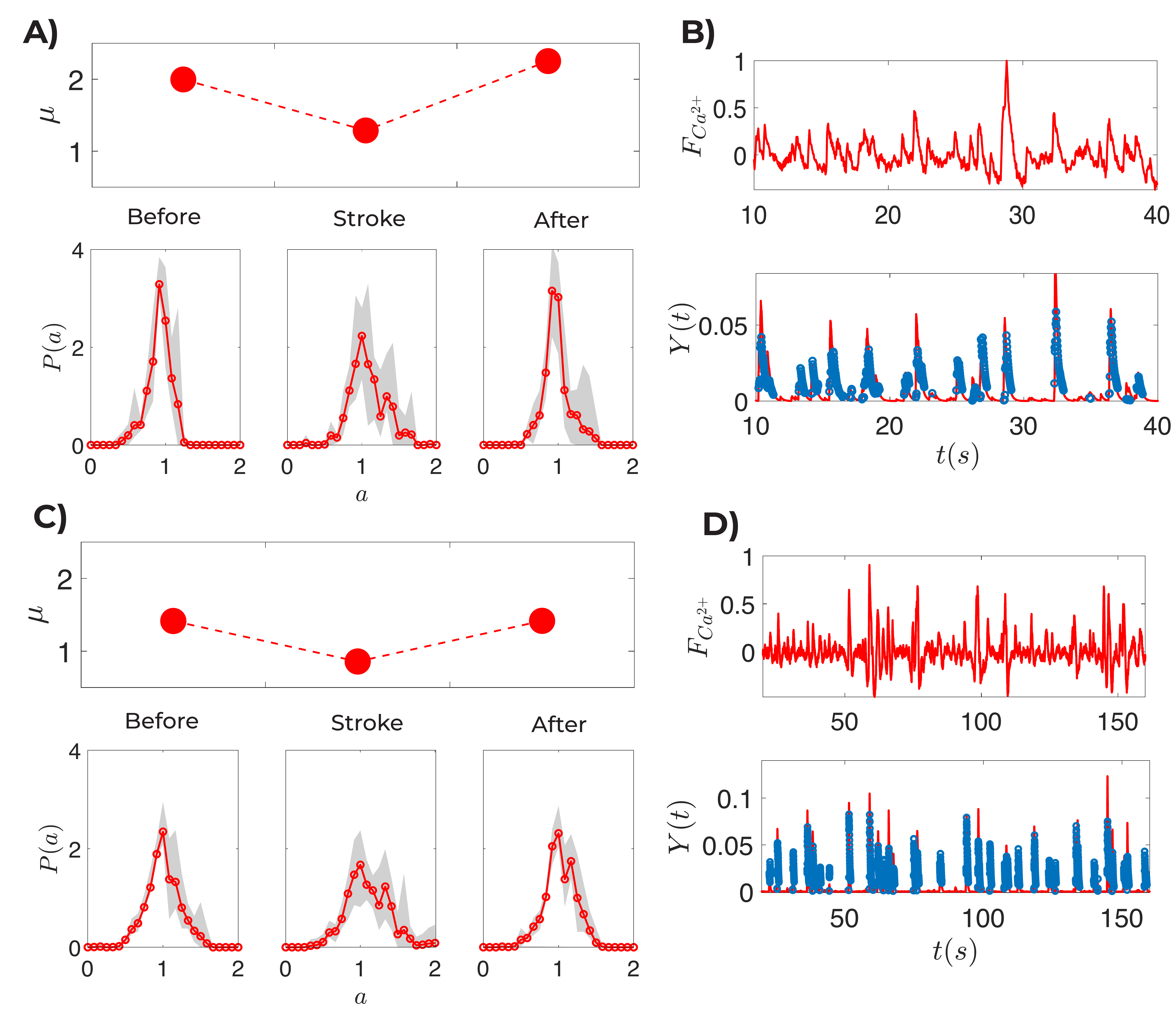}
    \caption{Panels A) and B) show results obtained from the resting data. In panel A) the distribution $P(a)$, averaged over 5 mice, is plotted for three successive windows of activity: before the stroke, immediately after stroke and three weeks after the stroke. In the reconstruction scheme a network with all--to--all connections in assumed (see SI for the combined reconstruction of $P(\tilde{k})$ and $P(a)$). Before the stroke the distribution is narrow and peaked around one, the threshold of instability. The fraction of effective neurons which fall in the rightmost tail is relatively modest. Immediately after the stroke, the degree of excitability is generally enhanced. Then, as time passes, the distribution $P(a)$ regains the original shape, by lowering the population of excited units. The grey shadow sets the level of individual variability, as quantified by the maximum and minimum excursions of the average curves collected for each mouse. The skewness $\mu$ of $P(a)$, for each of the explored windows, is plotted with filled red circle in the top inset of panel A). The tendency to recovery the initial configuration is quantitatively substantiated. In panel B) the raw fluorescence data, the global field $Y(t)$ (in red) and the fitted solution (blue circles) are respectively displayed. Panel C) and D) report the analogous quantities obtained by analyzing the data recorded during the trial sessions (see Methods). Again the tendency of the system to enhance and then lower the level of excitability is clearly demonstrated.}
    \label{fig:3}
\end{figure*}

The analysis of the experimental data can be broken up into two distinct parts. First we shall turn to analyzing the data collected under resting state. Then we will consider  the fluorescent signal recorded when the mice are motor training. In both cases, the raster plots are computed for each mouse and for different days of measurements. The reconstruction algorithm is then run according to the prescription detailed above so as to simultaneously recover the density distributions $P(\tilde{k})$ and $P(a)$ \footnote{Notice that the  coupling strength $g$, is set to the nominal value adopted in the forward simulations of the model,  see Fig. \ref{fig:1}. This latter value is experimentally justified and at present cannot be self-consistently generated as a byproduct of the inverse scheme. Remark however that the parameter $g$, could be in principle absorbed in the  definition of the rescaled degree $\tilde{k}$, a choice which   materialize in a rigid shift in the recovered distribution of connectivities.}. The results of this analysis are annexed as Supplementary Information: the $P(\tilde{k})$ profiles are consistently biased towards the rightmost edge
$\tilde{k}=k/N \rightarrow 1$, implying that the underlying network is densely connected. This observation is indeed in line with the computed raster plots: the firings of the fictitious neurons associated to each lattice point of the selected domain are highly synchronized, possibly implying a packed web of paired connections.  From a more fundamental point of view, and  as recalled above, each pixel integrates the signal from an extended pool of actual neurons. This translates into
a spatial resolution constraint which seems compatible with a local mean-field ansatz,  for an accurate interpretation of the data which transcends the detailed knowledge of connections between pairs of neurons.
Motivated by the results reported in the SI, we freeze therefore the network architecture to an all-to-all topology and set to reconstruct the a priori unknown $P(a)$, as the only source of heterogeneity under the HMF reductionist approach (see the setting analyzed in panels F,G,H of Fig. \ref{fig:1} for the benchmark application). The results of the analysis are displayed in Fig. \ref{fig:3} for both the resting and training settings. In both cases, and before the stroke, a peaked $P(a)$ profile is found with a modest fraction of (coarse-grained) neurons above the threshold of excitability. This is a reference working condition which appears quite reasonable under non pathological operating conditions. The distribution of the intensities gets distorted immediately after the stroke, the degree of  excitability being significantly enhanced. The effect is reproducible across the population of mice, as it can be appreciated by looking at the region colored in grey, which reflects the fluctuations due to individual variability. As time progresses during the rehabilitation training, the $P(a)$ converges back to its initial shape, an observation that can be made quantitative by monitoring the skewness of the reconstructed distributions (red circles in panels A and C of Fig. \ref{fig:3}). Hyperexcitability is indeed one of physiological mechanisms which are known to be triggered by the sudden death of a population of neurons, as e.g. following a stroke \cite{HyperExcitability1,HyperExcitability2,HyperExcitability3} . 
 
 Accordingly, hyperexcitable tissue (from the high baseline fluorescence) have been observed in the penumbra in the acute phase on the same mouse model, which progressively returned to lower levels in spontaneously recovering stroke mice \cite{AllegraMascaro}. Further studies show in fact that neurons in the penumbra region, the area adjacent to that where the stroke takes place, exhibit an increased firing rate, as the consequence of a reduced efficacity of $GABA_A$ synapses. This observation is consistent with the results that we have here reported, and which follows a direct implementation of the proposed inverse scheme.

\section*{Discussion and conclusion}

Recovering functional and structural information from in vivo measurements of brain activity is a challenge of broad applied and fundamental 
relevance. Working in this framework we have here proposed, and successfully  tested,  a novel reconstruction method which aims at inferring information   
on the distributions of both firing rates and networks connectivity, from global activity measurements. The technique builds on a variant of the celebrated  
Leaky-Integrate and Fire (LIF) model,  with short-term plasticity.  The heterogeneity in the assigned currents results in a non trivial, aperiodic, global field, 
which shares striking similarity with the homologous quantity, accessed in direct experiments. The proposed inverse protocol deals with an effective reductionist approach, which implements an apt version of the 
Heterogeneous Mean Field (HMF) ansatz: the dynamics of the system is organized in different classes, each associated with distinct currents $a$ and (rescaled) in-degree $\tilde{k}$. The reduced model is driven by the global 
field and the ensuing output used to fuel a self-consistent iterative scheme, which seeks at interpolating the field itself. When challenged against synthetic data, the inversion scheme 
is capable of returning an accurate description of the two distributions. Fluorescent microscopy data can be used to obtain experimentally tailored raster plots, which represent the 
ideal input for running the devised reconstruction machinery. The recovered distributions of intensities capture the phenomenon of  neuronal hyperexcitability, which has been experimentally found to interest the 
penumbra region, following a stroke event. This observation suggests that the proposed method could serve as a viable non-invasive strategy to track the   
recovery of pre-stroke brain functions.   As a general remark, it is worth stressing that, in many cases of interest, as in the present setting,
data from neural experiments integrate global synaptic signals, which extend typically over relatively wide areas of the brain.  The inverse procedure, based on the HMF recipe, 
moves from a coarse-grained representation of neuronal activity, from the scale of single neurons up to the pixels, which sets the image resolution of the recorded calcium waves.
The analysis that we have carried out indicates  that one can successfully frame the scrutinized dynamics in the context of simple microscopic models,  so as to interpret, and eventually reproduce, relevant 
macroscopic patterns at the brain scale. This implies in turn that a {\sl reductionist}, i.e. microscopically minimal, approach proves often adequate when aiming at 
describing the collective behavior of a complex neural system. By decorating the microscopic level with additional modeling ingredients
certainly allows one to grasp  a full load of specific features, which prove undoubtedly relevant when aiming at a characterizing punctual aspects of 
the individual neuronal activity. On the other hand, key collective effects, emerging from the sea of interacting units, seems to be modestly affected by such fine
details, which hence result in higher order corrections to basic interpretative frameworks.  For the case at hand, the imposed variability in the current $a_i$ enforces an effective modulation 
of the neural activity. {\sl A posteriori} we can argue that quenched disorder in the firing rates amounts to a simple heuristic recipe for taking 
into account  the effects of inhibitory neurons, which are responsible for tuning the excitatory neurons response. Stated differently, quenched disorder  surrogates the 
effect of the inhibitory components in the brain. Further extension of the method are in principle possible which enables one to resolve existing correlations between 
the topological and dynamical degrees of freedom.

\section*{Material and Methods}
  
\subsection{Experimental setup}
 
 All procedures involving mice were performed in accordance with the rules of the Italian Minister of Health, authorization n.183/2016-PR. Mice were housed in clear plastic cages under a 12 h light/dark cycle and were given ad libitum access to water and food. We used a transgenic mouseline (C57BL/6J-Tg(Thy1GCaMP6f)GP5.17Dkim/J)   (referred   to   as   GCaMP6f   mice) expressing a genetically-encoded fluorescent calcium indicator controlled by the Thy1 promoter. 

 {\bf Photothrombotic stroke induction}
All surgical procedures were performed under Isoflurane anesthesia (3 \% induction, 1.5 \% maintenence, in 1L/min oxygen). The animals were placed into a stereotaxic apparatus (Stoelting, Wheat Lane, Wood Dale, IL 60191) and, after removing the skin over the skull and the periosteum, the primary motor cortex (M1) was identified (stereotaxic coordinates +1,75 lateral, -0.5 rostral from bregma). Five minutes after intraperitoneal injection of Rose Bengal (0.2 ml, 10 mg/ml solution in Phosphate Buffer Saline (PBS); Sigma Aldrich, St. Louis, Missouri, USA), white light from an LED lamp (CL 6000 LED, Carl Zeiss Microscopy, Oberkochen, Germany) was focused with a 20X objective (EC Plan Neofluar NA 0.5, Carl Zeiss Microscopy, Oberkochen, Germany) and used to illuminate the M1 for 15 min to induce unilateral stroke in the right hemisphere. A cover glass and an aluminum head- post were attached to the skull using transparent dental cement (Super Bond, C\&S). Afterwards, the animals were placed in their cages until full recovery. 

{\bf  Motor training protocol on the M-Platform}
Mice were allowed to become accustomed to the apparatus before the first imaging session so that they became acquainted with the new environment. The animals were trained by means of the M- Platform, which is a robotic system that allows mice to perform a retraction movement of their left forelimb \cite{spalletti1,spalletti2, AllegraMascaro}. All the animals performed one week (5 sessions) of daily training before stroke induction and 4 weeks after the lesion in the M1.

{\bf  Wide-field fluorescence microscopy}
The custom-made wide-field imaging setup was equipped with a 505 nm LED (M505L3 Thorlabs, New Jersey, United States) light was deflected by a dichroic filter (DC FF 495-DI02 Semrock, Rochester, New York USA) on the objective (2.5x EC Plan Neofluar, NA 0.085, Carl Zeiss Microscopy, Oberkochen, Germany). The fluorescence signal was selected by a band pass filter (525/50 Semrock, Rochester, New York USA) and collected on the sensor of a high-speed complementary metal-oxide semiconductor (CMOS) camera (Orca Flash 4.0 Hamamatsu Photonics, NJ, USA). Images (100 $\times$ 100 pixels, pixel size 52 $\mu$m) were acquired at 25 Hz.

\section*{Acknowledgments}

This project received funding from the European Union's Horizon 2020 research and innovation programme under grant agreements No. 785907 (Human Brain Project) and 654148 (Laserlab-Europe), and from the EU programme H2020 EXCELLENT SCIENCE?European Research Council (ERC) under grant agreement ID No. 692943 (BrainBIT)

%\clearpage

%\clearpage

\appendix

\section{Testing the validity of the HMF approximations}

\begin{figure*}[h!]
    \centering
      \includegraphics[scale=0.3]{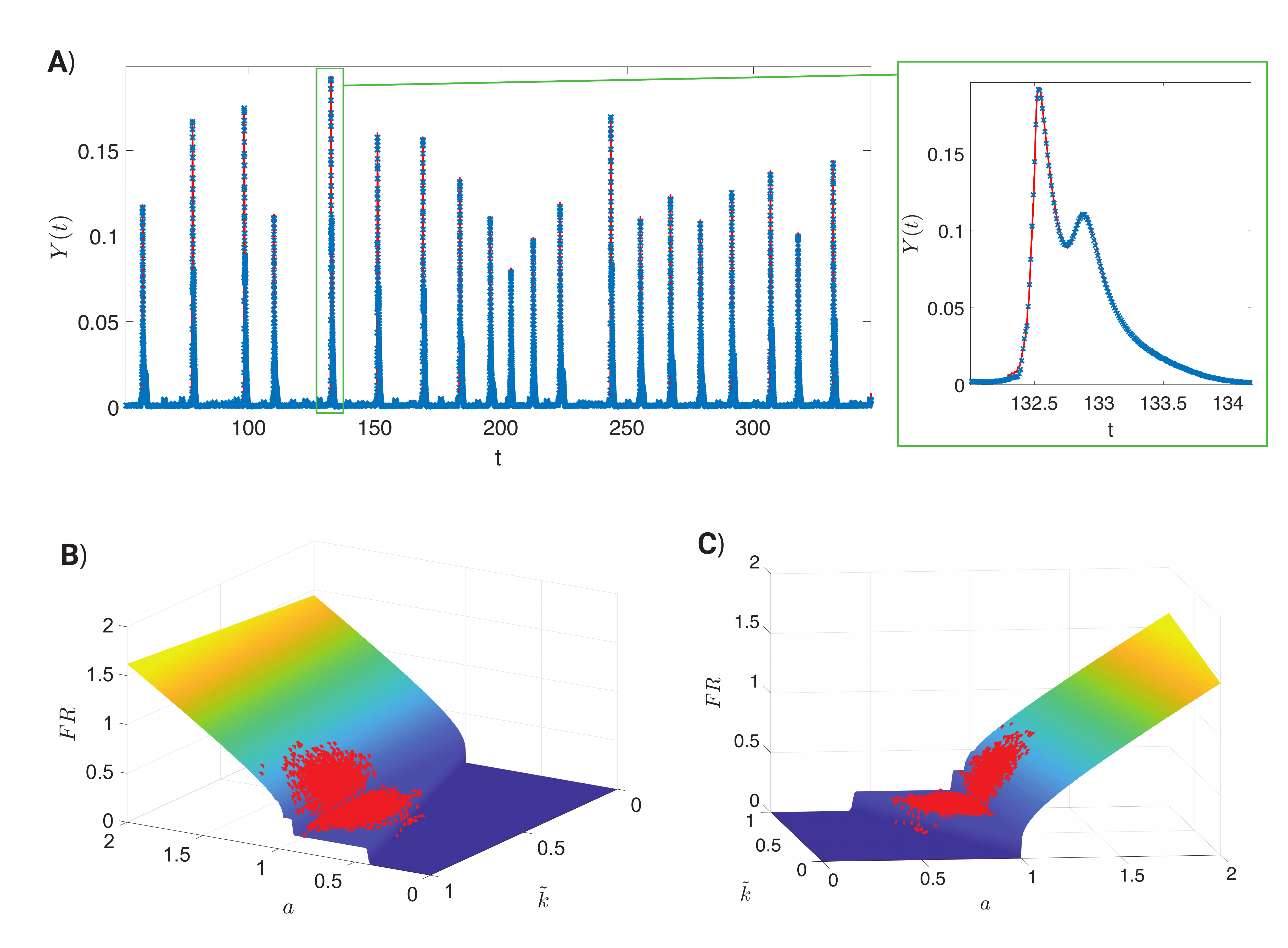}
     \caption{Panel A: the global field ($Y(t)$) as obtained from a direct integration of the LIF dynamics is shown with a solid red line. The reconstructed $\bar{Y}(t)$ signal, displayed with blue circles,  is obtained from the true distributions $P(\Tilde{k})$ and $P(a)$, by implementing the HMF reduction scheme. Here, $P(\Tilde{k})$ is a unimodal bell shaped distribution with $\langle \Tilde{k} \rangle = 0.7$, $\sigma_{\Tilde{k}}=0.082$. Similarly, $P(a)$  is characterized by $\langle a \rangle = 0.9$ and $\sigma_{a}=0.1$. Panel B: the Firing Rate ($FR$) is plotted as a function of $\Tilde{k}$ and $a$. The colored surface follows the HMF ansatz, while red crosses stem from direct integration of the original LIF model formulated in the space of the nodes.   Panel C is a different view of the results displayed in panel B.}
    \label{fig:S1}
\end{figure*}

The aim of this Section is to elaborate on the validity of the Hamiltonian Mean Field (HMF) approximation for the examined Leak Integrate and Fire (LIF) model of interacting neurons. To this end, we start by computing the global field $Y(t)$, from direct integration of the LIF equations for a given realization of the embedding network and assigned values of the currents. The distributions $P(\Tilde{k})$ and $P(a)$ are hence known a priori  and can be used to generate $\bar{Y}(t)$, an approximate representation of the global field, as follows Eq. (5) in the main body of the paper, and upon integration of the HMF dynamics, see  Eqs. (6-7). As it can be appreciated by visual inspection of panel A in Fig. \ref{fig:S1}
the field $\bar{Y}(t)$ obtained within the HMF scheme aligns almost perfectly with the signal which to be approximated. This in turn provides an a posteriori validation for the HMF working ansatz.

As a further check, we computed the individual Firing Rate ($FR$), i.e. the number of spikes per unit time, as a function of the $\Tilde{k}$ and $a$. The (red) stars plotted in panels B and C of Fig. \ref{fig:S1} refer to the $FR$ as obtained by analyzing the time series obtained by integrating of the LIF model in the space of the nodes. The stars lay  almost exactly on the manifold which is computed based on the HMF approximation. More precisely, for any given pair $(\Tilde{k},a)$ we calculate the corresponding $FR$, under the HMF assumption, yielding the surface depicted with an appropriate color code, this latter reflecting the estimated value of the $FR$ indicator. Panels B and C offer different views of the same plot. The agreement between predicted and measured quantities points again to the validity of the HMF interpretative framework.

\section{Further synthetic tests: reconstructing of a bi-modal $P(a)$}
\begin{figure*}[h!]
    \centering
      \includegraphics[scale=0.3]{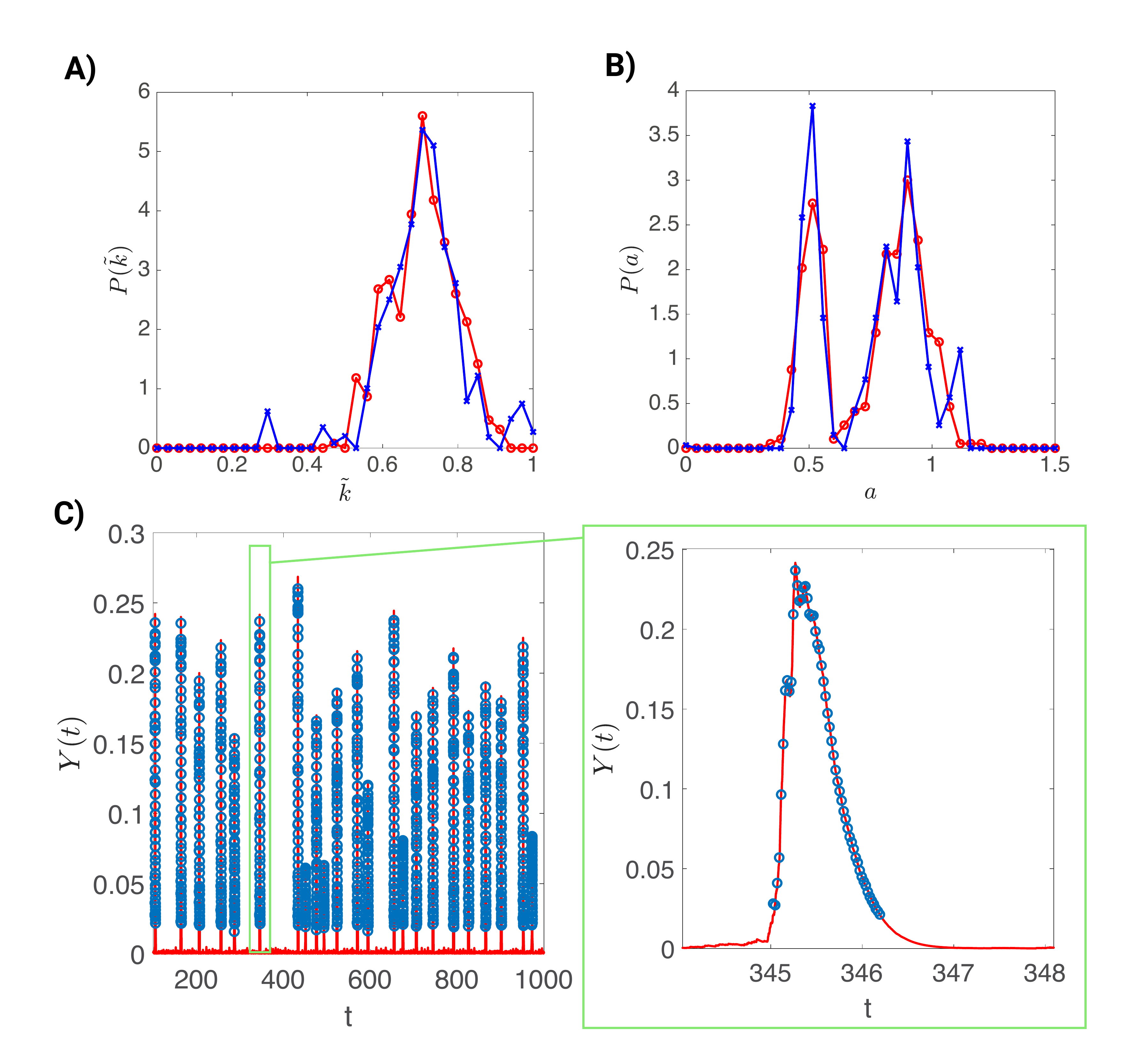}
     \caption{ Panel A: the true in-degree distribution $P(\Tilde{k})$ (here $\langle k \rangle = 0.7$, $\sigma_{\Tilde{k}}=0.082$) is shown in red (circles), while the recovered distribution is depicted in blue (crosses). Panel B: the true bi-modal distribution of the external currents $P(a)$ is shown in red (circle) and the recovered distribution in blue (crosses). Panel C: the true global field ($Y(t)$) is represented with a solid red line. Blue circles refer instead to the reconstructed signal $\bar{Y}(t)$.}
    \label{fig:S2}
\end{figure*}

To further test the robustness of the proposed reconstruction protocol, we challenge  the method against  simulated data obtained for a bimodal $P(a)$ distribution. The results of the tests, which follow the conceptual scheme outlined  in the main body of the paper, are displayed in Fig. \ref{fig:S2} and confirm the predictive ability of the proposed method. The distributions, $P(\Tilde{k})$ and $P(a)$, and the global field $Y(t)$ are nicely interpolated by their homologues reconstructed quantities.

\section{Reconstructing $P(\Tilde{k})$ and $P(a)$ from experimental data}

\begin{sidewaysfigure*}[ht!]
    \centering
      \includegraphics[scale=0.28]{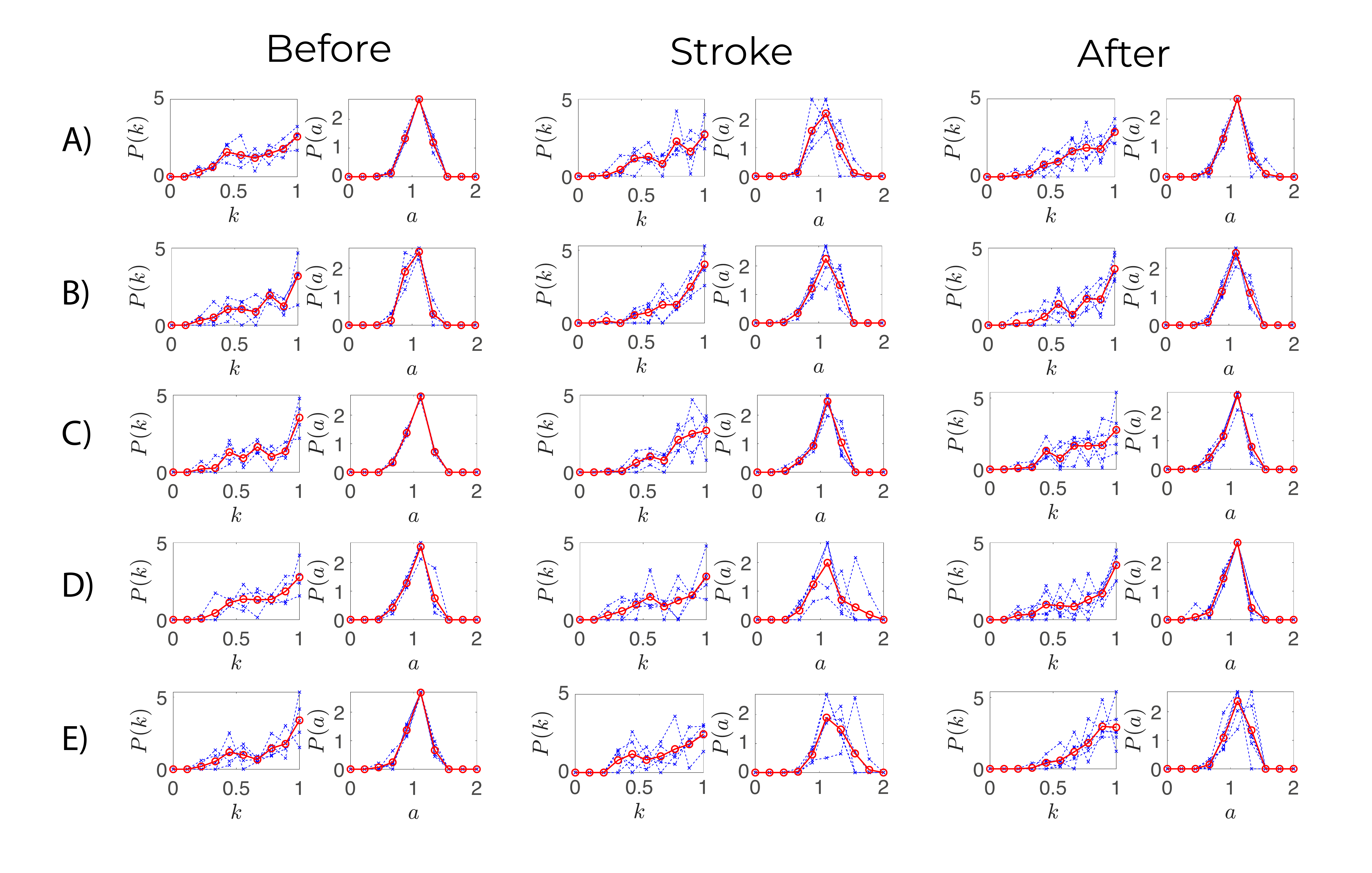}
     \caption{Recovered $P(\Tilde{k})$ and $P(a)$ for $5$ mice (identified by the letters running from A to E). Under the heading {\it Before}, we report the results obtained by analyzing data collected before the stroke over $4$ different days of observation. Blue light traces (crosses) refer to reconstructions obtained from data collected over one single day, among those belonging to the group of pertinence. Red curves (circles) stand for the distribution averaged over individual traces.  Under the heading {\it Stroke}, we report results which refer to $5$ days immediately after the stroke with an analogous organization of the displayed plots and employed symbols. Under the heading {\it After}, we have organized the results which follow the analysis of the physiological signals, as acquired three weeks after the stroke.}
    \label{fig:S3}
\end{sidewaysfigure*}

In this Section, we present the results obtained when reconstructing simultaneously the distribution $P(\Tilde{k})$ and $P(a)$ from experimental  data, in this case longitudinal wide-field fluorescence microscopy data of cortical functionality in groups of awake Thy1-GCaMP6f mice. For details that relate to the experiments and to the implemented reconstruction procedure, we refer to the main body of the paper. The results of the analysis are displayed in Fig. \ref{fig:S3} and organized for different mice (rows, from A to E) and physiological conditions (before/stroke/after, grouped in different macro-columns). Blue traces (crosses) refer to individual days of experiments, within the explored window. The red lines (circles)  stand for the average  of the reconstructed distributions. Only signals obtained during the training sessions are sufficiently long to allow for the simultaneous reconstruction of $P(\Tilde{k})$ and $P(a)$. In all cases, the recovered $P(\Tilde{k})$ attain their global maximum at $\Tilde{k}=1$. This justifies our assumption of an all-to-all arrangement of the underlying network which in turn allows us to focus on the recovery of $P(a)$ (see main paper).  However, the widening of the $P(a)$ immediately after the stroke, as discussed in the main paper, is also visible in the general setting when $P(\Tilde{k})$ and $P(a)$ are recovered at the same time.

%
%\clearpage


\begin{thebibliography}{99}


\bibitem{bialek} 
E. Schneidman, M. J. Berry II, R. Segev, W.  Bialek, Nature, {\bf 440}(7087), 1007 (2006).
\bibitem{cocco} 
S. Cocco, S. Leibler, R. Monasson,  Proceedings of the National Academy of Sciences, {\bf 106}(33), 14058-14062 (2009).
\bibitem{het_coding} 
S. J. Tripathy, K. Padmanabhan, R.C. Gerkin, N.N. Urban,  Proceedings of the National Academy of Sciences, {\bf 110}(20), 8248-8253 (2013).
\bibitem{deghani} 
N. Dehghani, A. Peyrache, B. Telenczuk, M. Le Van Quyen, E. Halgren, S. S. Cash, et al  Scientific reports {\bf 6}, 23176 (2016).
\bibitem{neuromod} 
L. K. Kaczmarek, I. B.  Levitan,  (Eds.) Neuromodulation: the biochemical control of neuronal excitability. Oxford University Press, USA (1987).
\bibitem{balance_epilepsy} 
R. W. Olsen, M. Avoli,  Epilepsia, {\bf 38} (4), 399-407 (1997).
\bibitem{spalletti1}	C. Spalletti et al., {\it Elife}, 6 (2017).
\bibitem{AllegraMascaro}    A.L. Allegra Mascaro et al., {\it Cell Rep},  {\bf 28}(13) 3474-3485 (2019).
\bibitem{diVolo1} 
R. Burioni, M. Casartelli, M. di Volo, R. Livi, A. Vezzani, Sci. Rep. {\bf 4}, 4336 (2014);
\bibitem{diVolo2}
M. di Volo, R. Burioni, M. Casartelli, R. Livi, A. Vezzani, Phys. Rev. {\bf E 90}, 022811 (2014);
\bibitem{diVolo3}
M. di Volo, R. Burioni, M. Casartelli, R. Livi, A. Vezzani, Phys. Rev. {\bf E 93}, 012305 (2016).
\bibitem{AFCI} 
I. Adam, D. Fanelli, T. Carletti and G. Innocenti, Eur. Phys. J. {\bf B 92}, 99 (2019).
\bibitem{HyperExcitability1} K. Schiene, C. Bruehl, K. Zilles, M. Qu, G. Hagemann,
M. Kraemer, and O. W. Witte, Journal of Cerebral Blood Flow and Metabolism {\bf 16}, 906 (1996).
\bibitem{HyperExcitability2} T. Neumann-Haefelin, G. Hagemann, and O. W. Witte,
Neuroscience letters {\bf 193}, 101 (1995).
\bibitem{HyperExcitability3} D. Berger, E. Varriable, L. Michiels van Kessenich, H.J. Hermann and L. de Arcangelis, Scientific Reports, to appear (2019).
\bibitem{TM}
M. Tsodyks and H.  Markram,  Proc. Natl. Acad. Sci. USA {\bf 94}, 719 (1997).
\bibitem{TPM}
M. Tsodyks, K. Pawelzik and H. Markram,  Neural Comput. {\bf 10}, 821 (1998).
\bibitem{DLLPT}
M. di Volo, R. Livi, S. Luccioli, A. Politi and  A. Torcini,  Phys. Rev. {\bf E 87}, 032801 (2013).
\bibitem{diVolo4}
M. di Volo and R. Livi, J. Chaos Solitons Fractals {\bf 57}, 54 (2013).
\bibitem{diVolo5}
M. di Volo, R. Livi, S. Luccioli, A. Politi and A. Torcini,  Phys. Rev. {\bf E 87}, 032801 (2013).
\bibitem{Olmi1}
S. Olmi, R. Livi, R., A. Politi and A. Torcini,  Phys. Rev. {\bf E 81}, 046119 (2010).
\bibitem{Olmi2}
S. Luccioli, S. Olmi, A. Politi, and A. Torcini, Phys. Rev. Lett. {\bf 109}, 138103 (2012).
\bibitem{Brunel}
M. Richardson, N. Brunel and V. Hakim,  J. Neurophysiol. {\bf 89}, 2538 (2003).
\bibitem{pittorino} F. Pittorino, M. Ibanez-Berganza, M. di Volo, A. Vezzani, and R. Burioni
Phys. Rev. Lett.  {\bf 118}  098102 (2017).
\bibitem{Chicchi} L. Chicchi, et al.  in preparation
\bibitem{spalletti2}	C. Spalletti et al.,  {\it Neurorehabil Neural Repair}, {\bf 28} (2) 188-96 (2014).

\end{thebibliography}
\end{document}